\documentstyle[osa,manuscript,epsf]{revtex}  % DON'T CHANGE
\begin{document}                % INITIALIZE - DONT CHANGE
\title{Neutron Electric Dipole Moment in SUSY SU(5) GUT}
\author{J.~Tabei\footnote{Email: jun@hep.phys.waseda.ac.jp}}
\address{
Department of Physics, Waseda University, Tokyo 169, Japan
}
\author{H.~Hotta\footnote{Email: hotta@hep.phys.waseda.ac.jp}}
\address{
Institute of Material Science and Technology, Waseda University, 
Tokyo 169, Japan
}
\maketitle

\begin{abstract}                % DON'T CHANGE THIS LINE
The existing estimation of the Electric Dipole Moment(EDM) 
of the neutron in the SUSY SU(5) GUT model was quite smaller 
than one in the MSSM, despite of its increased degree of freedom. 
This paradoxical result can be resolved 
by appropriately estimating the CP-violation in the Higgs sector. 
As a result, the neutron EDM in this model 
is estimated lager than the existing one as expected. 
\end{abstract}
\section{Introduction}               % Introduction goes below.
The Electric Dipole Moment(EDM) of an elementary particle 
is generated by the violation of CP \cite{He}. 
In the standard model \cite{He},\cite{Czarnecki},  
the neutron EDM ($d_n$) is theoretically evaluated as: 
\begin{equation}
d_n = 10^{-31} \sim 10^{-34}\mbox{(e cm)}
\ . \label{eq:sme}
\end{equation}
On the other hand, 
the experimental upper limit on the neutron EDM is: 
\begin{equation}
|d_n| < 6.3 \times 10^{-26} \mbox{(e cm)} 
\ , \label{eq:ul}
\end{equation} 
at present\cite{Harris}.  
However, this experimental limit will be improved 
in the near future. 
In several extensions of the standard model, 
it is well-known that the neutron EDM 
is expected or evaluated to be larger than 
one in the original standard model. 
Thus, once the neutron EDM is observed larger than 
the prediction Eq.(\ref{eq:sme}) in the standard model, 
the reality of the other models is increased. 
In this paper, we estimate the neutron EDM in the 
SUSY SU(5) GUT model as one of the extensions of the 
standard model. 
Prior to the authors, 
the neutron EDM in this model is already 
evaluated by A. Romanino and A. Strumia \cite{Romanino}. 
However, in their existing estimation\cite{Romanino}, 
there is not so much difference of the neutron EDMs 
between in the SUSY SU(5) GUT model and in the standard model 
except the large $\tan\beta$ settings, 
because the CP violating phase only in the quark sector of 
the mass matrix is taken into account, 
and such phase in the Higgs sector is essentially neglected. 
Generally speaking, according to the Kobayashi-Maskawa 
theorem \cite{KM}, a mass matrix consisted of $3 \times 3$ 
or more components can include the complex phase 
to violate CP. 
Thus, we introduce the complex phase into 
the Higgs sector of the mass matrix, 
because the Higgs/Higgsino superfields have five 
components in this model, and the existence of this 
new phase is one of the features of the SUSY SU(5) GUT model 
in contrast to the standard model or MSSM. 
When trying to estimate the neutron EDM 
in the SUSY SU(5) GUT model, this new mechanism to violate CP 
should be taken into consideration. 
In the following sections, all the estimations 
are carried out with including this new mechanism 
to violate CP symmetry.

% SECTION 2
\section{Estimation of Neutron EDM}
In this section, we present a sequence of the derivation 
from the general theory of SUSY SU(5) GUT to the estimation 
of the neutron EDM. 
\subsection{Superpotential}
In the SUSY SU(5) GUT model, the superpotential is given 
as follows \cite{Nilles},\cite{Georgi}: 
\begin{eqnarray}
  V_{S} &=& \lambda_{1} 
             \left(
                \frac{1}{3}
                 \Sigma^{a}_{b}
                 \Sigma^{b}_{c}
                 \Sigma^{c}_{a}
              + \frac{1}{2}
                 m
                 \Sigma^{a}_{b}
                 \Sigma^{b}_{a}\
             \right)
%\nonumber\\
\label{eq:sp}\\
          && +\lambda_{2} \bar{H}_{a} 
             \left(
                \Sigma^{a}_{b}
              + 3 m' \delta^{a}_{b}
             \right)
            H^{b}
\nonumber\\
          && + f_{j k} 
            \varepsilon_{u v w a b}
            H^{u}
            X^{v w}_{j}
            X^{a b}_{k}
\nonumber\\
         && + g_{j k} 
            \bar{H}_{a}
            X^{a b}_{j}
            {Y}_{k b}
            \nonumber\\
            && +h.c.
            \nonumber
            \ , 
\end{eqnarray}
where, $a$ and $b$ are SU(5)'s, $j$ and $k$ are 
flavors' indices, respectively. 
On the Higgs supermultiplets $H$ and $\bar{H}$, 
$H^{a}$ is a $5$-representation and $\bar{H}_{a}$ 
is a $\bar{5}$-representation. 
On the matter supermultiplets $X$ and $Y$, 
$X^{a b}_j$ are the representation of $10$'s, and 
$Y_{j a}$ are the $\bar{5}$'s. 
Note that $X^{a b}_j=-X^{b a}_j$. \\ 
\ \\
\noindent
The complex phase parameter $\theta$ to violate CP 
will be introduced into the second term of the superpotential 
Eq.(\ref{eq:sp}) as the Higgs sector: 
\begin{equation}
3 \lambda_{2} m'\bar{H} H \: \longrightarrow 
\: 3 \lambda_{2} m'\bar{H}e^{i \theta} H 
\ . \label{eq:cpb}
\end{equation}
For more detail on the phase introduction, see the section 3. 

\subsection{How to break SUSY SU(5) down}
It is well-known that most of the symmetries in this model, 
like SUSY, are broken in our real world; 
the sequence to break the symmetries down is as follows: 
\begin{enumerate}
\item The original SU(5) symmetry is breaking to 
SU(3)$\times$SU(2)$\times$U(1) with the energy scale 
getting lower by the Higgs mechanism of the heavy Higgs 
bosons in 24-representation. 
\item Moreover, the supersymmetry is broken spontaneously by 
added the soft breaking terms\cite{Nilles} as: 
\begin{equation}
m^{2}_{0} \phi^{\ast}_{i} \phi_{i}
+(m_{0}A_{0}W_{3}+B_{0} \mu_{0}h_{1}h_{2}+h.c.)
+m_{1/2}\bar{\lambda}_{j} \lambda_{j} 
\ , \label{eq:soft}
\end{equation}
where, $\phi$'s are the scalar fields, 
$W_{3}$ stands for the trilinear scalar terms, 
$h_{1}$ and $h_{2}$ are the two Higgs multiplets, 
and $\lambda$'s are the gauge fermions. 
$m_{0}$ is the universal scalar mass, $A_{0}$ is the 
universal trilinear 
coupling constant, $B_{0}$ is the universal 
bilinear coupling constant, 
and $m_{1/2}$ is the universal gaugino mass. 
On the meanings or the values of these introduced parameters 
are discussed in the section 3. 
\item Furthermore, SU(2) symmetry is spontaneously 
breaking to the electromagnetic gauge symmetry as usual: 
\begin{equation}
\mbox{SU(2)} \times \mbox{U(1)} 
\rightarrow \mbox{U(1)}_{\mbox{em}} \ . 
\end{equation}
The observations of the neutron EDM are hold in this energy region. 
\end{enumerate}

\subsection{EDM formulae}
In order to estimate the EDM of a quark current, 
the shapes of one-loop graphs shown in Fig.1 are regarded. 
The representations of each particles coincide with 
their physical states, where the explicit forms of 
the representations are shown in Appendix B. 
The diagrams including the virtual loop of the chargino, 
neutralino or the gluino can generate the CP-odd contributions 
to the current amplitude. Therefore, the contributions of 
these three kinds of the diagrams are taken into account. 
The graphs including the virtual loop of the heavy gauginos
({\em i.e.}, X-ino or Y-ino) and the heavy Higgsinos are also 
possible to contribute CP-oddly, however, they are neglected. 
Roughly speaking, since the EDMs are proportional 
to the inverse of the virtual particle masses ; the EDM 
values of such graphs including heavy virtual particles are 
extremely small compared to the graphs of the charginos, 
neutralinos, or the gluino. 

\ \\
For the purpose of calculating the loop correction to 
a current of 
the quark with spin 1/2, 
the current amplitude is decomposed according to the 
form factors\cite{} as follows: 
\begin{eqnarray}
&&\left<
q_{finit}(p')|j^{\mu} (q)|q_{initial}(p)
\right>
\label{eq:ampj}\\
&&=\bar{q}_{finit}(p') \left[
  \gamma^{\mu} F_{1} (q^{2})+i \sigma^{\mu \nu} q_{\nu}
\frac{F_{2} (q^{2})}{2m}
+ \gamma_{5} \sigma^{\mu \nu} q_{\nu} \frac{F_3(q^{2})}{2m}
\right.
\nonumber \\
&& \mbox{\hspace{35mm}}
\left. +\left( \frac{q^{2}}{2m}\gamma^{\mu}-q^{\mu} \right) \gamma_{5} 
F_{A}(q^{2})
+ \cdots
\right]q_{initial}(p)
\ , \nonumber
\end{eqnarray}
where, $j_{\mu}(q)$ is the current with 
four-momentum transfer $q=p'-p$.
$F_3(q^2)$ is the CP-odd form factor. \\
\ \\
The EDM($d_f$) of a quark with the flavor $f$ is given by 
\begin{equation}
d_{f}=- \frac{e}{2m} F_{3}(q^2 \rightarrow 0)
\ . \label{eq:edm}
\end{equation}
\noindent
By making use of this definition, the neutron EDM($d_n$) is 
written as: 

\begin{equation}
d_{n}=\frac{4}{3} d_{d} - \frac{1}{3} d_{u} 
\ , \label{eq:edmd}
\end{equation}
where, $d_d$ means the d-quark EDM, as $d_u$ is the u-quark EDM. 
\\
\noindent
Moreover, the Chromoelectric Dipole Moment(CDM)
\cite{He}, \cite{SUSY}is also taken into account. 
The CDM is defined as the factor $d^{CDM}$ 
of the effective operator arise in the QCD Laglangian: 

\begin{equation}
{\cal L}_{CDM}=
-\frac{i}{2}d^{CDM}\bar{q}\sigma_{\mu\nu}\gamma_{5}T^{a}
qG^{\mu\nu a}
\ , 
\end{equation}
where, $T^{a}$'s stand for the generators of SU(3). \\ 
The CDM contributions to the neutron EDM\cite{He} is as follows: 
\begin{equation} 
d_{n}=\frac{1}{3}e\left( 
\frac{4}{3} d^{CDM}_{d} + \frac{2}{3} d^{CDM}_{u} \right)
\ . \label{eq:edmc}
\end{equation}
For more details on the loop calculations of the EDM and 
CDM formulae, see Appendix C.

% SECTION 3
\section{Evaluation of parameters}
In order to evaluate the soft breaking mass parameters, 
on the correction of the renormalization group flow 
at one loop level is discussed in this section. 
The explicit forms of the Renormalization Group Equations(RGEs) 
are shown in Appendix A. 
In order to solve the RGEs, we assume the following 
conditions.

\begin{itemize}
\item[i)]The gauge coupling constants are bent at 
$M_{SUSY} \simeq 1$(TeV), 
and unified at 
$M_{SU(5)} \simeq 10^{16}$(GeV)\cite{Amaldi}. 

\item[ii)]Yukawa coupling constants of the top and bottom quarks 
are given to coincide with their values 
at electroweak scale $M_{Z}$ independent of $\tan{\beta}$. 

\item[iii)] The spontaneous SUSY breaking down scale $M_{X}$ is as: 
\begin{eqnarray}
M_X=G^{-1/4}_{F}M^{1/2}_{P}\sim 6.0\times 10^{10}\mbox{(GeV)}\ , 
\end{eqnarray}
where, $G_{F}$ is the Fermi coupling constant and 
$M_{P}$ is the Planck mass scale. 
Therefore, $M_{X}$ is 
fixed to $6.0\times 10^{10}$(GeV). 
Note that this condition is {\em ad-hoc} for 
the pure SUSY SU(5) GUT, 
because this is a consequence of the super Higgs effect 
in hidden sector based on N=1 supergravity grand unified 
theory\cite{Nilles}. Also this is mere a formal boundary condition, 
indeed the values of $m_{\tilde{q}}$ 
and $A_{\tilde{q}}$ are almost independent of $M_{X}$. 

\item[iv)] All the masses of the supersymmetric scalar particles 
are set equal to $m_{0}$ at energy scale $M_{X}$. 

\item[v)] $B_{0}=(A_{0}-1)m_{0}$ is assumed. This is 
another {\em ad-hoc} condition for the pure SUSY SU(5) GUT 
adopted from the minimal supergravity theories\cite{Nilles}. 

\item[vi)] Once an assumed $B_{0}$ at $M_{X}$ is given, 
the parameters like $m_{h1}$ or $m_{h2}$(mass parameter of 
two Higgs doublet) at $M_{Z}$ scale can be obtained 
by solving the RGEs directly. On the other hand, to break 
the electro-weak symmetry down as usual, two conditions: 
\begin{eqnarray}
2B\mu&=&-(m^{2}_{h1}+m^{2}_{h2}+2\mu^{2})\sin{2\beta}
\ , \\
\mu^{2}&=&-\frac{m^2_{Z}}{2}+\frac{m^{2}_{h1}-m^{2}_{h2}\tan^2{\beta}}
{\tan^2{\beta}-1}
\ ,
\end{eqnarray}
are required. As the results of these two equations, 
$B$ and $\mu$ at $M_{Z}$ are derived, however, this value of 
$B$ at $M_{Z}$ should be coincident with another estimation. 
To match both two $B$ values at $M_{Z}$, the initial $B_0$ 
at $M_X$ is tuned recursively and decided by numerical analysis. 
Thus, $B_{0}$ ( and also $A_{0}$) is fixed. 

\item[vii)] $\tan\beta$ is restricted as $2<\tan\beta<40$, 
to avoid emerging the Landau pole divergence of Yukawa coupling 
constant of the top quark. 

\item[viii)] The other parameters are cited from the last report 
of Particle Data Group\cite{PDG}. 
\end{itemize}
\ \\
\noindent
As it is slight confusing because there exist many introduced 
parameters with mass dimension, we give a summary of them 
in the Table 1. 
\ \\
\noindent
With these conditions, several additional restrictions 
on other parameters are obtained from the RGE analysis. 
The existence of $m_{1/2}$ in Eq.(\ref{eq:soft}) implies 
the universality of all the gaugino masses at the SUSY breaking 
energy scale $M_{X} = 6.0 \times 10^{10}$(GeV). The condition 
$M_{SU(5)}>M_X$ renders this universality, because, 
the unification of the gauge coupling constants is broken 
at $M_{SU(5)}$ in the first, however, the supersymmetry 
to make the gauginos massless still remains until the energy $M_X$. 
As the results, 
the following three relations between $m_{1/2}$ and each 
gaugino masses or gauge couplings are realized at one loop 
level\cite{Martin}: 
\begin{equation}
\frac{m_{i}(M_{Z})}{\alpha_{i}(M_{Z})}
=\frac{m_{1/2}}{\alpha_{i}(M_{X})} \ , \ (i=1, 2, 3)
\ , \label{eq:gaugino}
\end{equation}
where $\alpha_{i}$'s are the gauge coupling constants. 
Note that these relations give each (different) 
gaugino masses at $M_Z$ to solve the RGEs, however, they 
never mean the universality among the gauge couplings 
at $M_Z$ or $M_X$, 
{\em i.e.}, $\alpha_{i} \neq \alpha_{j} (i \neq j)$, in general.\\ 
Furthermore, several constraints are derived 
among $\lambda_{2}$, $m'$, and the CP-violating phase $\theta'$ 
in the Higgs sector. 
The parameter $\mu$ can be regarded as the Higgsino mass, 
and is defined as: 
\begin{equation}
\mu=\lambda_{2} \left( 3m-3m'e^{i\theta} \right) 
\ \ . \label{eq:mu}
\end{equation}
By our RGE analysis, $|\mu|$ is about 1(TeV). 
As $m$ implies the VEV of the Higgs bosons 
in the $24$-representation, 
thus the order of $m$ is $10^{16}$(GeV). 
The unified mass of the 5 and $\bar{5}$'s SU(3) 
heavy Higgs bosons ($m_{color}$) is given as: 
\begin{equation}
m_{color}=\lambda_{2}\left( 2m+3m'e^{i\theta}\right) 
\ \ . \label{eq:mc}
\end{equation}
Quite small contributions to $m_{color}$ from 
the soft breaking terms Eq.(\ref{eq:soft}) are neglected. 
The order of the mass $m_{color}$ is estimated 
as $10^{16}$(GeV) by the proton decay analysis\cite{PD}. 
Due to these two relations Eq.(\ref{eq:mu}) and Eq.(\ref{eq:mc}), 
the phase $\theta$ is restricted as: 
\begin{equation}
|\sin\theta|
\leq
\sin\theta_{max}
=
\frac{5|m_{color}||\mu|}{3|m_{color}|^{2}+2|\mu|^{2}}
\sim
O(10^{-13}) 
\ \ , \label{eq:tmax}
\end{equation}
and the ratio $m'/m$ is restricted as: 
\begin{equation}
\frac{3|m_{color}|-2|\mu|}{3(|m_{color}|+|\mu|)}
\leq
\frac{m'}{m}
\leq
\frac{3|m_{color}|+2|\mu|}{3(|m_{color}|-|\mu|)}
\ \ , 
\end{equation}
with the quantitative relation between $|m_{color}|$ and $|\mu|$, 
\begin{equation}
|m_{color}| \gg |\mu|
\ \ .
\end{equation}
Therefore, 
\begin{equation}
m'/m \simeq 1
\ \ , 
\end{equation}
is valid in good approximation. This result coincides with 
the fact that $m'$ is 
introduced as the fine-tuning parameter to keep the Higgsino 
mass in the usual region $\sim$ 1(TeV), {\em i.e.}, the order 
of $m'$ should be the same with $m$ as $10^{16}$(GeV). 
On the other hand, the parameter $\lambda_{2}$ is written as: 
\begin{equation}
\lambda_{2}
=\frac{|m_{color}|}{|2m+3m'e^{i\theta}|} 
\ , \label{eq:L2}
\end{equation}
therefore, the order of the parameter $\lambda_{2}$ 
is evaluated as $O(10^{-1})$. 
When $\mu$ is rewritten as: 
\begin{equation}
\mu = |\mu|\exp{(i\theta')} \ , 
\end{equation}
the complex phase $\theta'$ of $\mu$ is given by: 
\begin{equation}
\sin{\theta'}
=
\frac{3\lambda_{2}}{|\mu|}m'|\sin{\theta}|
\ . \label{eq:sint2}
\end{equation}
The phase $\theta'$ of $\mu$ as the CP-violating phase 
in the Higgs sector is given by this function of $|\mu|$, 
$\lambda_{2}$, $m'$ and $\sin{\theta}$. Obviously, 
the maximum $\sin{\theta}$ as $O(10^{-13})$ corresponds 
to the maximum CP-violating phase $\theta'$. The order of the 
maximum $\sin\theta'$ is derived from Eq.(\ref{eq:sint2}) as: 
\begin{equation}
|\sin{\theta}| 
\Rightarrow 
\sin{\theta_{max}} \ , \ 
\sin{\theta'}
\simeq 1
\ , \ 
\end{equation}
by making use of two conditions Eq.(\ref{eq:tmax}) 
and Eq.(\ref{eq:L2}).  
In Fig.2, the phase dependence of the neutron EDM 
is given as a function of $\sin{\theta}$. 
In the following sections, $\theta=\theta_{max}$ is fixed. 
\noindent

% SECTION 4
\section{Numerical Analysis and Results}
\noindent
The rest free parameters are $m_{1/2}$, $m_0$, 
and $\tan{\beta}$. 
These parameters are expected to settle in such regions 
as usual: 
\begin{itemize}
\item
$m_2$ :  TeV order.
\item
$m_0$ :  less than a few TeV .
\item
$\tan{\beta}$ :  2 $\sim$ 40
($\simeq |m_{t}|/|m_{b}|$ at $M_Z$)
\end{itemize}
\ \\
\noindent
In Fig.3, the neutron EDMs are shown 
as functions of $m_{1/2}$ : $1\leq m_{1/2} \leq 10$(TeV) 
for $\tan \beta$=3, 5, and 10. $m_{0}=$1(TeV) is fixed. 
The flat line is the experimental upper limit of EDM. 
Obviously, $m_{1/2}$'s value is required to be 
larger than 4.5(TeV) by the experimental limit. \\
\ \\
\noindent
In Fig.4, on the $m_{1/2}$ dependence of the neutron EDM 
is drawn for $m_{0}=$1, 3, and 5(TeV). $\tan \beta=5$ is fixed. 
The flat line means the experimental upper limit as in Fig.3, 
$m_{1/2}$ is necessary to be larger than 5(TeV). The curves 
imply that the neutron EDM is not sensitive 
on $m_{0}$. \\
\ \\
\noindent
Figure 5 summarizes the topographic plots of EDM's experimental 
upper limit as functions of $m_{0}$ and $m_{1/2}$ 
for $\tan\beta$=3, 5, and 10. 
These plots show the two facts - $m_{1/2}$'s value should be 
larger than a few TeV, and $\tan\beta$ is required to be small.
The EDM is more sensitive on $m_{1/2}$ 
rather than $m_{0}$. 
%\pagebreak

% SECTION 5
\section{Conclusion and Summary}
\noindent
In summary, 
the neutron EDM in the SUSY SU(5) GUT model is 
obtained by introducing the complex phase into 
the mass matrix of the Higgs sector. As the results, 
the neutron EDM of this mechanism is derived 
significantly larger than existing estimations by the CP 
violation in the quark sector. According to this results, 
the constraints on the $m_{0}$, $m_{1/2}$ and $\tan\beta$ 
become more strict than existing evaluations, however,  
the allowed parameter region still remains. 
Additionally, we should take care of the phases of other 
parameters, like $m_{1/2}$'s, because such phases can 
cancel\cite{Brhlik} the effect of the introduced phase 
into the Higgs sector. If such phases were possible to exist, 
the evaluation of the EDM in this paper should be 
reconsidered as the maximum, {\em i.e.}, the strictest estimation. 
Anyway, once the neutron EDM would be observed larger 
than the prediction of the standard model 
Eq.(\ref{eq:sme}), the other models beyond the standard model, 
including the SUSY SU(5) GUTs(predicting large EDM), 
will be more realistic. 
\\

\section*{Acknowledgements}
The authors thank to Dr. N. Oshimo for many 
useful instructions on the 
supersymmetric theories. We also appreciate 
Profs. I. Ohba and H. Nakazato 
for a lot of advice. \\

%\pagebreak

\appendix
%Appendix A
\section{Equations of Renormalization Group}
\setcounter{equation}{0}
\renewcommand{\theequation}{A\arabic{equation}}
\noindent
The RGEs at one loop level are as follows\cite{Martin},\cite{RGE}: 
\begin{eqnarray}
\frac{d\alpha_{a}}{dt}&=&-2b_{a}\alpha^2_{a}
\ , \\
\frac{dm_{a}}{dt}&=&-2b_{a}\alpha_{a}m_{a}
\ , \\
\frac{d\alpha_{t}}{dt}&=&2\alpha_{t}(-c^{(1)}_{a}\alpha_{a}+6\alpha_{t}
+\alpha_{b})
\ , \\
\frac{d\alpha_{b}}{dt}&=&2\alpha_{b}(-c^{(2)}_{a}\alpha_{a}
+\alpha_{t}+6\alpha_{b})
\ , \\
\frac{dA_{\tilde{t}}}{dt}&=&-2c^{(1)}_{a}\alpha_{a}m_{a}/m_{0}
+12\alpha_{t}A_{\tilde{t}}+2\alpha_{b}A_{\tilde{b}}
\ , \\
\frac{dA_{\tilde{u},\tilde{c}}}{dt}&=&-2c^{(1)}_{a}\alpha_{a}m_{a}/m_{0}
+6\alpha_{t}A_{\tilde{t}}
\ , \\
\frac{dA_{\tilde{b}}}{dt}&=&-2c^{(2)}_{a}\alpha_{a}m_{a}/m_{0}
+2\alpha_{t}A_{\tilde{t}}+12\alpha_{b}A_{\tilde{b}}
\ , \\
\frac{dA_{\tilde{d},\tilde{s}}}{dt}
&=&-2c^{(2)}_{a}\alpha_{a}m_{a}/m_{0}
+6\alpha_{b}A_{\tilde{b}}
\ , \\
\frac{dB}{dt}&=&-2c^{(3)}_{a}\alpha_{a}m_{a}
+6\alpha_{t}m_{0}A_{\tilde{t}}+6\alpha_{b}m_{0}A_{\tilde{b}}
\ , \\
\frac{dm^2_{h1}}{dt}&=&-2c^{(3)}_{a}\alpha_{a}m^{2}_{a}
+6\alpha_{b}\Sigma^2_{b}
\ , \\
\frac{dm^2_{h2}}{dt}&=&-2c^{(3)}_{a}\alpha_{a}m^{2}_{a}
+6\alpha_{t}\Sigma^2_{t}
\ , \\
\frac{dm^2_{\tilde{u},\tilde{c},\tilde{d},\tilde{s}L}}{dt}&=&
-2c^{(4)}_{a}\alpha_{a}m^2_{a}
+\frac{1}{5}\alpha_{1}\mbox{Tr}(Ym^2)
\ , \\
\frac{dm^2_{\tilde{u},\tilde{c}R}}{dt}&=&-2c^{(5)}_{a}
\alpha_{a}m^2_{a}
-\frac{4}{5}\alpha_{1}\mbox{Tr}(Ym^2)
\ , \\
\frac{dm^2_{\tilde{d},\tilde{s}R}}{dt}&=&-2c^{(6)}_{a}\alpha_{a}m^2_{a}
+\frac{2}{5}\alpha_{1}\mbox{Tr}(Ym^2)
\ , \\
\frac{dm^2_{\tilde{t},\tilde{b}L}}{dt}&=&2\alpha_{t}\Sigma^2_{t}
+2\alpha_{b}\Sigma^2_{b}
-2c^{(4)}_{a}\alpha_{a}m^2_{a}
+\frac{1}{5}\alpha_{1}\mbox{Tr}(Ym^2)
\ , \\
\frac{dm^2_{\tilde{t}R}}{dt}&=&4\alpha_{t}\Sigma^2_{t}
-2c^{(5)}_{a}\alpha_{a}m^2_{a}
-\frac{4}{5}\alpha_{1}\mbox{Tr}(Ym^2)
\ , \\
\frac{dm^2_{\tilde{b}R}}{dt}&=&4\alpha_{b}\Sigma^2_{b}
-2c^{(6)}_{a}\alpha_{a}m^2_{a}
+\frac{2}{5}\alpha_{1}\mbox{Tr}(Ym^2)
\ , 
\end{eqnarray}
where, the dimensionless energy scale variable $t$ is defined 
as $t=(1/4\pi)\ln(Q/M_{SU(5)})$ derived from the bare energy 
scale variable $Q$. 
$m_{a}$'s are the gaugino masses, 
and $\alpha_{a}=g^2_{a}/4\pi$'s are 
the gauge coupling constants. 
$A_{\tilde{q}}$ and $m_{\tilde{q}}$ are squarks' 
trilinear coupling constants and their masses, respectively. 
The suffix $L$ or $R$ indicates each partner's chirality. 
The suffix $a$ is always summed up from 1 to 3. 
The optional coefficients and variables are defined as: 
\begin{eqnarray}
b_{a}&=&\left\{
\begin{array}{ll}
(-41/10,19/6,7) & \qquad, \ M_{Z}\le Q \le M_{SUSY}\\ 
(-33/5,-1,3) & \qquad, \ M_{SUSY} \le Q \le M_{SU(5)} 
\end{array} \right. 
\ , \\
c^{(1)}_{a}&=&(13/15,3,16/3)
\ , \\
c^{(2)}_{a}&=&(7/15,3,16/3)
\ , \\
c^{(3)}_{a}&=&(3/5,3,0)
\ , \\
c^{(4)}_{a}&=&(1/15,3,16/3)
\ , \\
c^{(5)}_{a}&=&(16/15,0,16/3)
\ , \\
c^{(6)}_{a}&=&(4/15,0,16/3)
\ , \\
\Sigma^2_{t}&=&(A^2_{\tilde{t}}m^2_{0}+m^2_{h2}
+m^2_{\tilde{t}L}+m^2_{\tilde{t}R})
\ , \\
\Sigma^2_{b}&=&(A^2_{\tilde{b}}m^2_{0}+m^2_{h1}
+m^2_{\tilde{b}L}+m^2_{\tilde{b}R})
\ . 
\end{eqnarray}

\noindent
All the Yukawa couplings except of the top and bottom quarks are 
neglected, because their values are too small to be taken 
into account. 
The exceptional two (relatively large) Yukawa couplings 
$\alpha_{t,b}=g^2_{t,b}/4\pi$ 
are defined as usual: 
\begin{eqnarray}
g_{t}
=\frac{g_{2}}{\sqrt{2}}\frac{m_{t}}{m_{W}}\frac{1}{\sin{\beta}}
\ , \\
g_{b}
=\frac{g_{2}}{\sqrt{2}}\frac{m_{b}}{m_{W}}\frac{1}{\cos{\beta}}
\ . 
\end{eqnarray}

\noindent
Furthermore, keeping the traceless conditions of the SU(5) 
gauge: 
\begin{equation}
\mbox{Tr}(Ym^2)=m^2_{0}\mbox{Tr}(Y)=0
\ , 
\end{equation}
where, the $5 \times 5$ diagonal matrix $Y$ is defined as: 
\begin{eqnarray}
Y=diag(-2/3,-2/3,-2/3,1,1)\ , 
\end{eqnarray}
are required to avoid the gravitational mixed anomaly\cite{Martin}. 
\ \\
\noindent
Since the initial or boundary conditions are necessary 
to solve the RGEs, they are given at the electro-weak scale 
$M_{Z} \simeq 0.9 \times 10^{2}$(GeV) for the gauge and 
Yukawa couplings, 
and at $M_{X} \simeq 6.0 \times 10^{10}$(GeV) scale for 
the other variables. 
Note that we set all the initial values of the soft breaking 
mass parameters equal to $m_{0}$, and an ad-hoc 
constraint $B_{0}=(A_{0}-1)m_{0}$ from the minimal supergravity 
theory is applied. 

%Appendix B
\section{Representations of particle states}
\setcounter{equation}{0}
\renewcommand{\theequation}{B\arabic{equation}}
As one of the general features of the quantum theory, 
the mixture among the particles with the same quantum numbers 
usually occurs, then, some non-diagonal elements 
arise in the mass matrix of bare particles. 
In order to coincide their representations with the physical 
particles, the mass matrix should be diagonalized and realized 
as possible as they can be. \\

\noindent
The gauginos and Higgsinos are mixed as the charginos and 
the neutralinos. As the results of the mixture, the mass matrix 
of the bare charginos is :
\begin{eqnarray}
\mbox{\boldmath $M_{C}$}=
\left( \begin{array}{cc}
  m_{2} & \sqrt{2} m_{W}\sin{\beta} \\
   \sqrt{2}m_{W} \cos{\beta} & \lambda_{2}b
  \end{array} \right)
  \ , \label{eq:cha} 
\end{eqnarray}
and of the bare neutralinos is: 
\begin{eqnarray}
\mbox{\boldmath $M_{N}$}=
\left( \begin{array}{l}
    \mbox{\hspace{18mm}}
  m_{1} \mbox{\hspace{5mm}} 0  
    \mbox{\hspace{13mm}}
   -m_{Z}\sin{\theta_{W}}\cos{\beta} 
    \mbox{\hspace{5mm}}
    m_{Z}\sin{\theta_{W}}\sin{\beta} \\
    \mbox{\hspace{20mm}}
  0 \mbox{\hspace{5mm}} m_{2} 
    \mbox{\hspace{13mm}}
    m_{Z}\cos{\theta_{W}}\cos{\beta} 
    \mbox{\hspace{5mm}}
   -m_{Z}\cos{\theta_{W}}\sin{\beta} \\
  -m_{Z}\sin{\theta_{W}}\cos{\beta}
   \mbox{\hspace{5mm}}
   m_{Z}\cos{\theta_{W}}\cos{\beta} 
    \mbox{\hspace{12mm}}
    0 \mbox{\hspace{3mm}} -\lambda_{2}b \\
   m_{Z}\sin{\theta_{W}}\sin{\beta}
   \mbox{\hspace{5mm}}
  -m_{Z}\cos{\theta_{W}}\sin{\beta} 
    \mbox{\hspace{10mm}}
    -\lambda_{2}b \mbox{\hspace{5mm}} 0 \\
  \end{array} \right) \ , 
\end{eqnarray}
\noindent
where, $m_{1}$ and $m_{2}$ stand for the gaugino masses, 
$m_{W}$ and $m_{Z}$ are 
W and Z boson masses, respectively. 
$b=3m-3m'e^{i\theta}$; $m$ is the Vacuum 
Expectation Value(VEV) of heavy Higgs bosons in 
the 24-representation. 
$\tan{\beta}$ is defined as $\tan{\beta}=v_{2}/v_{1}$; 
$v_{1}$ and $v_{2}$ are the VEVs 
of the Higgs scalar fields $\bar{5}$ and $5$, respectively.
These mass matrices are diagonalized by the biunitary 
transformation for the charginos \cite{Guion} as:
\begin{equation}
\mbox{\boldmath $U^{\ast}M_{C}V$}=
diag(|m_{C_{1}}|,|m_{C_{2}}|)
\ , 
\end{equation}
where, the suffices $C_{i}(i=1,2)$ represent the physical chargino 
states. Moreover, the neutralinos are diagonalized by 
the complex orthogonal transformation as: 
\begin{equation}
\mbox{\boldmath $N^{T}M_{N}N$}=
diag(|m_{N_{1}}|,|m_{N_{2}}|,|m_{N_{3}}|,|m_{N_{4}}|)
\ , 
\end{equation}
where, the suffices $N_{i}(i=1\sim 4)$ imply 
the physical neutralino states. 
Note that the representations of these diagonalized physical states 
as the charginos and neutralinos are used all through this paper, 
instead of the bare fields of the Higgsino and gaugino.
It is not necessary for the gluino to diagonalize the mass matrix 
in contrast to the charginos or neutralinos. 
Therefore, the gluino mass $m_{\tilde{g}}$ is simply equivalent 
to $m_{3}$ in Eq.(\ref{eq:gaugino}). 
We introduce a symbol $\tilde{g}$ as the gluino. \\

\noindent
The bare mass matrix ($6 \times 6$) of the u-type squarks
is given as: 
\begin{equation}
\mbox{\boldmath $M_{\tilde{U}}$}^{2}
=
\left( \begin{array}{cc} 
\mbox{\boldmath $m$}^{2}_{\tilde{q}L}+D_{L}
\mbox{\boldmath $1$}+\mbox{\boldmath $m$}^{2}_{U} &
\mbox{\boldmath $m$}_{U}(\mbox{\boldmath $A$}^{\ast}_{\tilde{q}}m_{0}
+\lambda_{2}b(v_{1}/v_{2}))\\
\mbox{\boldmath $m$}_{U}(\mbox{\boldmath $A$}_{\tilde{q}}m_{0}
+\lambda_{2} b^{\ast} (v_{1}/v_{2}))&
\mbox{\boldmath $m$}^{2}_{\tilde{q}R}+D_{R}
\mbox{\boldmath $1$}+\mbox{\boldmath $m$}^{2}_{U}
\end{array} \right) \ , 
\end{equation}
and of the d-type squarks is: 
\begin{equation}
\mbox{\boldmath $M_{\tilde{D}}$}^{2}
=
\left( \begin{array}{cc} 
\mbox{\boldmath $m$}^{2}_{\tilde{q}L}+D_{L}
\mbox{\boldmath $1$}+\mbox{\boldmath $m$}^{2}_{D} &
\mbox{\boldmath $m$}_{D}(\mbox{\boldmath $A$}^{\ast}_{\tilde{q}}m_{0}
+\lambda_{2}b(v_{2}/v_{1}))\\
\mbox{\boldmath $m$}_{D}(\mbox{\boldmath $A$}_{\tilde{q}}m_{0}+
\lambda_{2} b^{\ast} (v_{2}/v_{1}))&
\mbox{\boldmath $m$}^{2}_{\tilde{q}R}+D_{R}
\mbox{\boldmath $1$}+\mbox{\boldmath $m$}^{2}_{D}
\end{array} \right) \ , 
\end{equation}
where, 
\begin{eqnarray}
D_{L}
&=&
m^{2}_{Z}(T_{3U,D}-Q_{U,D} \sin^{2}{\theta_{W}})\cos{2\beta}\ , \\
D_{R}
&=&
m^{2}_{Z}Q_{U,D} \sin^{2}{\theta_{W}}\cos{2\beta} \ . 
\end{eqnarray}
$Q_{U,D}$ stands for the electric charge of the u,d-type quark, 
respectively. $T_{3U}=+1/2$, and $T_{3D}=-1/2$. 
$ \mbox{\boldmath $m$}_{U,D}$ denotes the u,d-type ordinal quark 
mass matrix$(3 \times 3)$, respectively. 
$\mbox{\boldmath $m$}_{\tilde{q}L}$ and 
$\mbox{\boldmath $m$}_{\tilde{q}R}$ are 
the matrices of the soft breaking mass parameters of the squarks. 
$\mbox{\boldmath $A$}_{\tilde{q}}$ is 
the trilinear coupling constants matrix. 
Both of the squark mass matrices$(6 \times 6)$ are diagonalized as: 
\begin{eqnarray}
\mbox{\boldmath $D^{\dagger}_{\tilde{u}}M_{\tilde{U}}D_{\tilde{u}}$}
&=&
diag(
|m_{\tilde{u}_{11}}|,|m_{\tilde{u}_{12}}|,|m_{\tilde{u}_{13}}|,
|m_{\tilde{u}_{21}}|,|m_{\tilde{u}_{22}}|,|m_{\tilde{u}_{23}}|
)\ , \\
\mbox{\boldmath $D^{\dagger}_{\tilde{d}}M_{\tilde{D}}D_{\tilde{d}}$}
&=&
diag(
|m_{\tilde{d}_{11}}|,|m_{\tilde{d}_{12}}|,|m_{\tilde{d}_{13}}|,
|m_{\tilde{d}_{21}}|,|m_{\tilde{d}_{22}}|,|m_{\tilde{d}_{23}}|
)\ , 
\end{eqnarray}
where, the suffices $\tilde{u},(\tilde{d})_{k,j}(k=1,2)$ mean 
the physical states of the u,(d)-type squarks 
of the $j$-th generation, respectively(the sub-index $k$ 
stems from two chiralities of the ordinal quarks as 
the superpartners of the squarks, however, this has 
no physical meaning). 
These definitions of the squark states are used 
all through this paper, like of the charginos and neutralinos. 
Note that we neglect all the non-diagonal 
elements of $\mbox{\boldmath $A$}_{\tilde{q}}$, 
$\mbox{\boldmath $m$}_{\tilde{q}L,R}$ 
and $ \mbox{\boldmath $m$}_{U,D}$ and their additional phase 
in the squark mass matrices to violate CP, however, this 
additional effect has already estimated by 
others\cite{Romanino},\cite{Inui}; the EDM generated by this 
additional phase is quantitatively quite smaller than one by the 
phase of the Higgs sector in our discussions.

%Appendix C
\section{Neutron EDM Formulae}
\setcounter{equation}{0}
\renewcommand{\theequation}{C\arabic{equation}}
The neutron EDM and CDM are given by following 
equations\cite{SUSY},\cite{Inui}. All the notations of the 
transformation matrices and their elements, like $V^{\ast}_{i2}$
in Eq.(\ref{eq:gam1}), are the same with Appendix B. 
The chargino contribution is: 
\begin{eqnarray}
d^{C}_{f}/e&=&-\frac{\alpha_{em}}{4\pi\sin^2{\theta_{W}}}
\sum_{k=1}^2 \sum_{i=1}^2 \mbox{Im}(\Gamma^{C}_{fik})
\frac{m_{Ci}}{m^2_{\tilde{f^{\prime}}k}}\nonumber\\
&&\times 
\left[
Q_{\tilde{f^{\prime}}}
B \left( \frac{m^2_{Ci}}{m^2_{\tilde{f^{\prime}}k}} \right)
+(Q_{f}-Q_{\tilde{f^{\prime}}})
A \left( \frac{m^2_{Ci}}{m^2_{\tilde{f^{\prime}}k}} \right) 
\right] 
\ , \\
d^{C-CDM}_{f}&=&-\frac{\alpha_{em}g_{3}}{4\pi\sin^2{\theta_{W}}}
\sum_{k=1}^2 \sum_{i=1}^2 \mbox{Im}(\Gamma^{C}_{fik})
\frac{m_{Ci}}{m^2_{\tilde{f^{\prime}}k}}
B \left( \frac{m^2_{Ci}}{m^2_{\tilde{f^{\prime}}k}} \right)
\ , 
\end{eqnarray}
where, $f=u,d$ for $f^{\prime}=d,u$, respectively. $\alpha_{em}$ 
is the fine-structure constant, $g_{3}=\sqrt{4\pi\alpha_{3}}$, 
and $\theta_{W}$ is the Winberg angle. 
Each definitions of two introduced functions $A(r)$ and $B(r)$ 
are: 
\begin{eqnarray}
A(r)=\frac{1}{2(1-r)^2}\left(3-r+\frac{2\ln{r}}{1-r}\right) 
\ , \\
B(r)=\frac{1}{2(1-r)^2}\left(1+r+\frac{2r\ln{r}}{1-r}\right) 
\ , 
\end{eqnarray}
respectively. The $\Gamma$'s are defined as: 
\begin{eqnarray}
\Gamma^C_{uik}&=&\kappa_{u}V^{\ast}_{i2}D_{d1k}
(U^{\ast}_{i1}D^{\ast}_{d1k}
-\kappa_{d}U^{\ast}_{i2}D^{\ast}_{d2k})
\ , \label{eq:gam1} \\
\Gamma^C_{dik}&=&\kappa_{d}U^{\ast}_{i2}D_{u1k}
(V^{\ast}_{i1}D^{\ast}_{u1k}
-\kappa_{u}V^{\ast}_{i2}D^{\ast}_{u2k})
\ , \label{eq:gam2} 
\end{eqnarray}
and 
\begin{eqnarray}
\kappa_{u}&=&\frac{m_{u}}{\sqrt{2}m_{W}\sin{\beta}}
\ , \\
\kappa_{d}&=&\frac{m_{d}}{\sqrt{2}m_{W}\cos{\beta}}
\ . 
\end{eqnarray}
The neutralino contribution is: 
\begin{eqnarray}
d^{N}_{f}/e
&=&\frac{\alpha_{em}}{4\pi\sin^2{\theta_{W}}}
\sum_{k=1}^2 \sum_{i=1}^4 \mbox{Im}(\Gamma^{N}_{fik})
\frac{m_{Ni}}{m^2_{\tilde{f}k}}
Q_{\tilde{f}}
B \left( \frac{m^2_{Ni}}{m^2_{\tilde{f}k}} \right)
\ , \\
d^{N-CDM}_{f}
&=&\frac{\alpha_{em}g_{3}}{4\pi\sin^2{\theta_{W}}}
\sum_{k=1}^2 \sum_{i=1}^4 \mbox{Im}(\Gamma^{N}_{fik})
\frac{m_{Ni}}{m^2_{\tilde{f}k}}
B \left( \frac{m^2_{Ni}}{m^2_{\tilde{f}k}} \right)
\ , \\
\Gamma^{N}_{fik}
&=&[-\sqrt{2}\{\tan{\theta_{W}}(Q_{f}-T_{3f})N_{1i}
+T_{3f}N_{2i}\}
D^{\ast}_{f1k}+\kappa_{f}N_{bi}D^{\ast}_{f2k}]\nonumber\\
&&\times
(\sqrt{2}\tan{\theta_{W}}Q_{f}N_{1i}D_{f2k}-\kappa_{f}N_{bi}D_{f1k})
\ , 
\end{eqnarray}
where, the suffix $b=3(4)$ for $T_{3f}=-1/2(+1/2)$, 
respectively. \\
The gluino contribution is: 
\begin{eqnarray}
d^{\tilde{g}}_{f}/e&=&-\frac{2\alpha_{3}}{3\pi}
\sum_{k=1}^2  \mbox{Im}(\Gamma^{\tilde{g}}_{fk})
\frac{m_{\tilde{g}}}{m^2_{\tilde{f}k}}
Q_{\tilde{f}}
B \left( \frac{m_{\tilde{g}}}{m^2_{\tilde{f}k}} \right)
\ , \\
d^{\tilde{g}-CDM}_{f}&=&\frac{\alpha_{3}g_{3}}{4\pi}
\sum_{k=1}^2  \mbox{Im}(\Gamma^{\tilde{g}}_{fk})
\frac{m_{\tilde{g}}}{m^2_{\tilde{f}k}}
C \left( \frac{m_{\tilde{g}}}{m^2_{\tilde{f}k}} \right)
\end{eqnarray}
where, 
\begin{eqnarray}
\Gamma^{\tilde{g}}_{fk}&=&D_{f2k}D^{\ast}_{f1k}
\ , \\
C(r)&=&\frac{1}{6(r-1)^2}
\left( 10r-26+\frac{2r\ln{r}}{1-r}-
\frac{18\ln{r}}{1-r} \right) \ . 
\end{eqnarray}
Note that all the generations of the squark 
are summed up in the virtual loops. 
At last, in order to compare with the EDM of the 
real neutron, the total sum of above EDM and CDM contributions 
is composed subject to the formulae Eq.(\ref{eq:edmd}) 
and Eq.(\ref{eq:edmc}).

%
% ({\it REVTEX} 3.0 automatically issues
% a \newpage command when the \begin{table} or \begin{figure}
% commands are used, so the figures and tables will be placed
% on separate pages by {\it REVTEX}).

% \begin{figure}  % Please send figures with disk, or separately if
%% if it is an e-mail submission. (Good photo or India ink drawing.)
% \caption{Please place your figure caption here.}
% \end{figure}

\begin{figure}
%\vspace{6cm}
% the next line size is for the submission only 
   \epsfxsize=15cm
% the next line size is for the publication 
%   \epsfxsize=8.5cm
   \centerline{\epsfbox{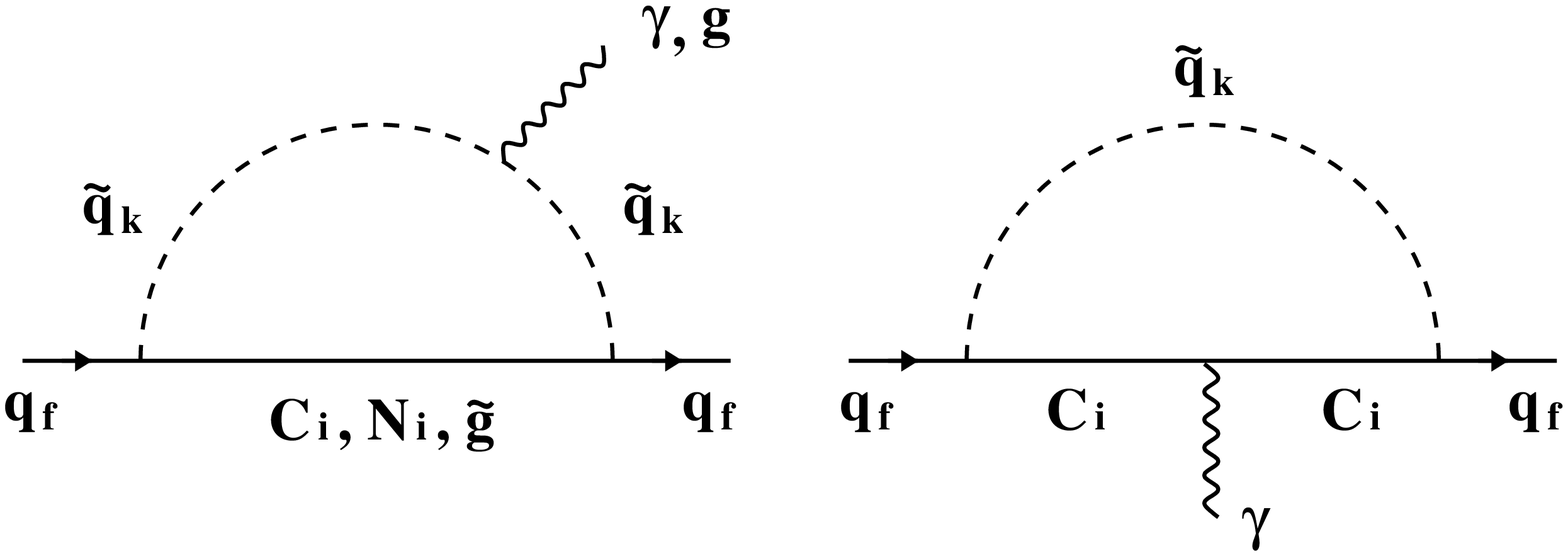}}
% for the preprint server
\caption{
Diagrams to contribute to EDM
}
\end{figure} 
\pagebreak

\begin{figure}
%\vspace{6cm}
% the next line size is for the submission only
   \epsfxsize=15cm
% the next line size is for the publication
%   \epsfxsize=8.5cm
   \centerline{\epsfbox{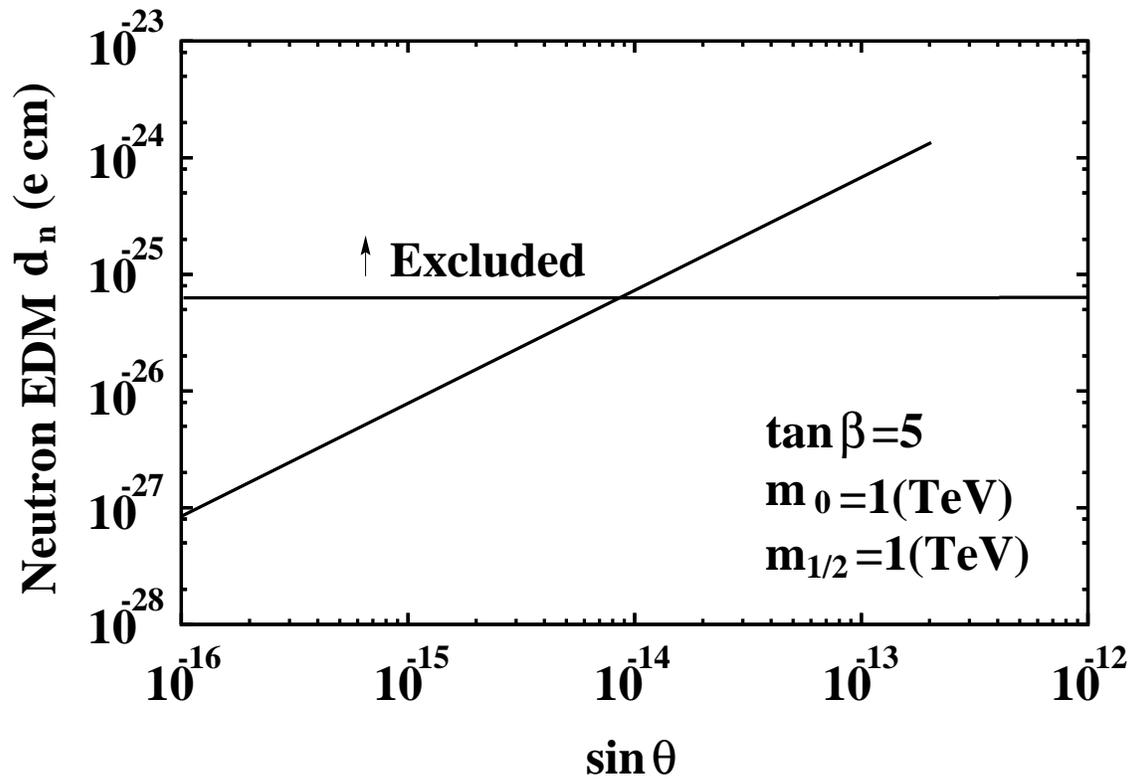}}
% for the preprint server
%\vspace{6cm}
\caption{
Neutron EDM as a function of $\sin \theta$
}
\end{figure}
\pagebreak

\begin{figure}
%\vspace{6cm}
% the next line size is for the submission only
   \epsfxsize=15cm
% the next line size is for the publication
%   \epsfxsize=8.5cm
   \centerline{\epsfbox{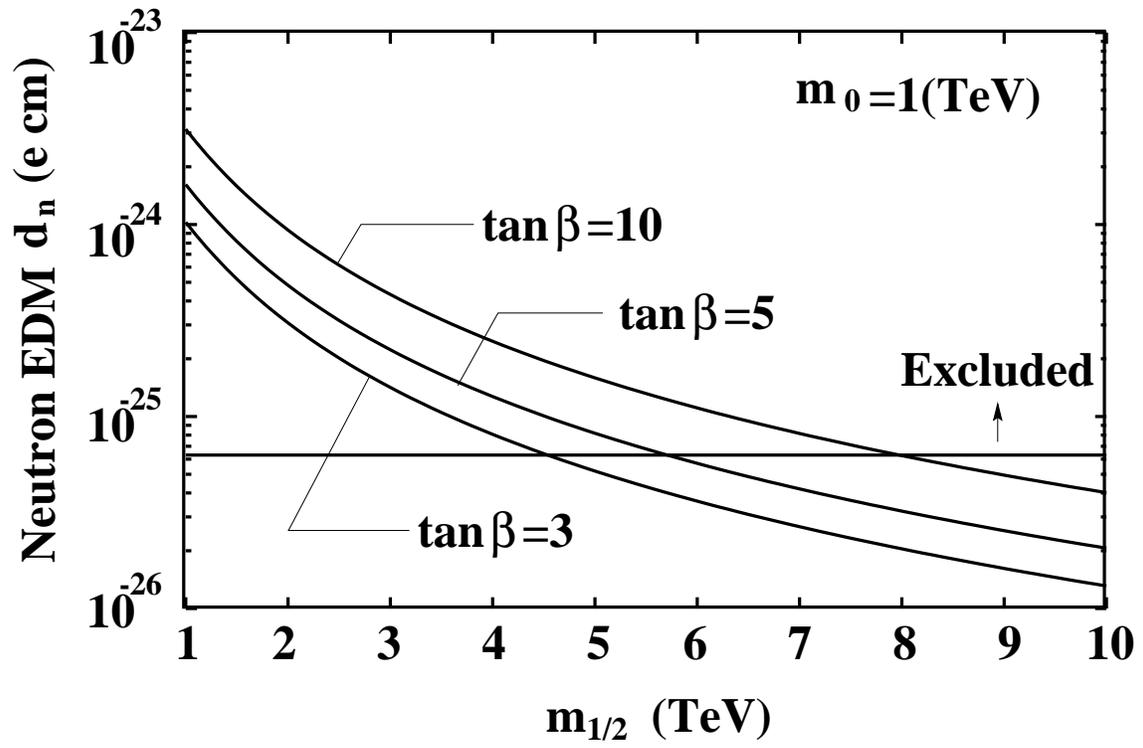}}
% for the preprint server
%\vspace{6cm}
\caption{
Neutron EDMs as functions of $m_{1/2}$ for 
several $\tan \beta$
}
\end{figure}
\pagebreak

\begin{figure}
%\vspace{6cm}
% the next line size is for the submission only
   \epsfxsize=15cm
% the next line size is for the publication
%   \epsfxsize=8.5cm
   \centerline{\epsfbox{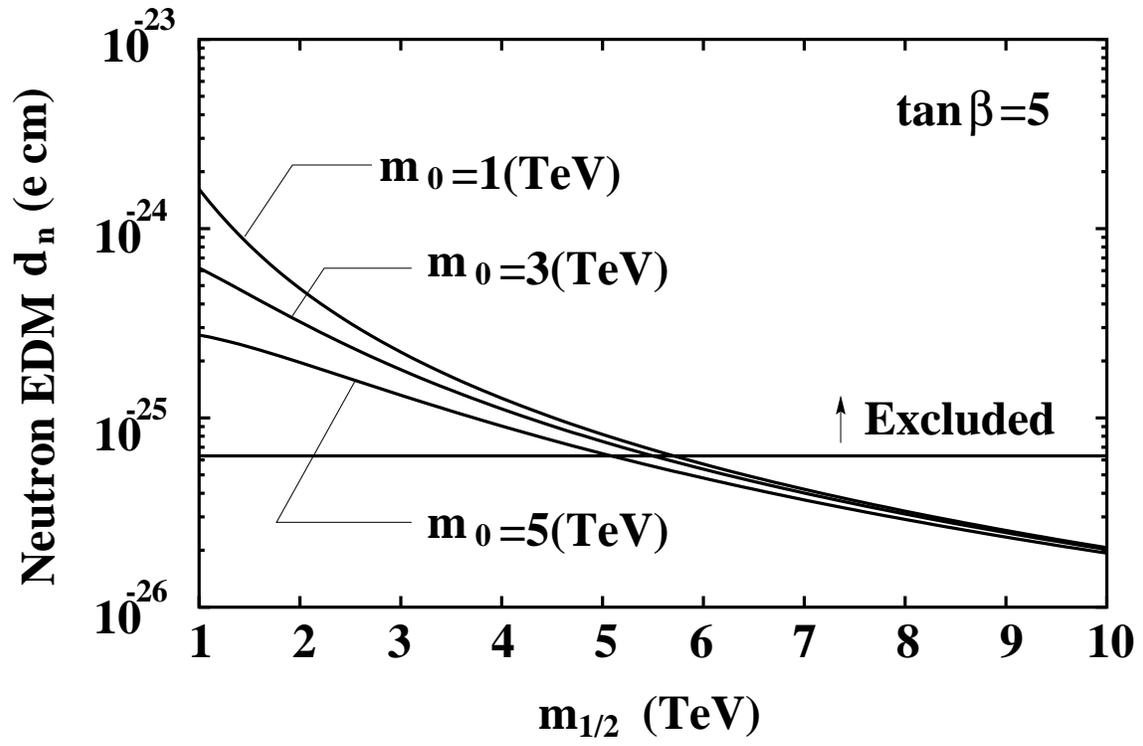}}
% for the preprint server
%\vspace{6cm}
\caption{
Neutron EDMs as functions of $m_{1/2}$ for 
several $m_{0}$
}
\end{figure} 
\pagebreak

\begin{figure}
%\vspace{6cm}
% the next line size is for the submission only
   \epsfxsize=15cm
% the next line size is for the publication
%   \epsfxsize=8.5cm
   \centerline{\epsfbox{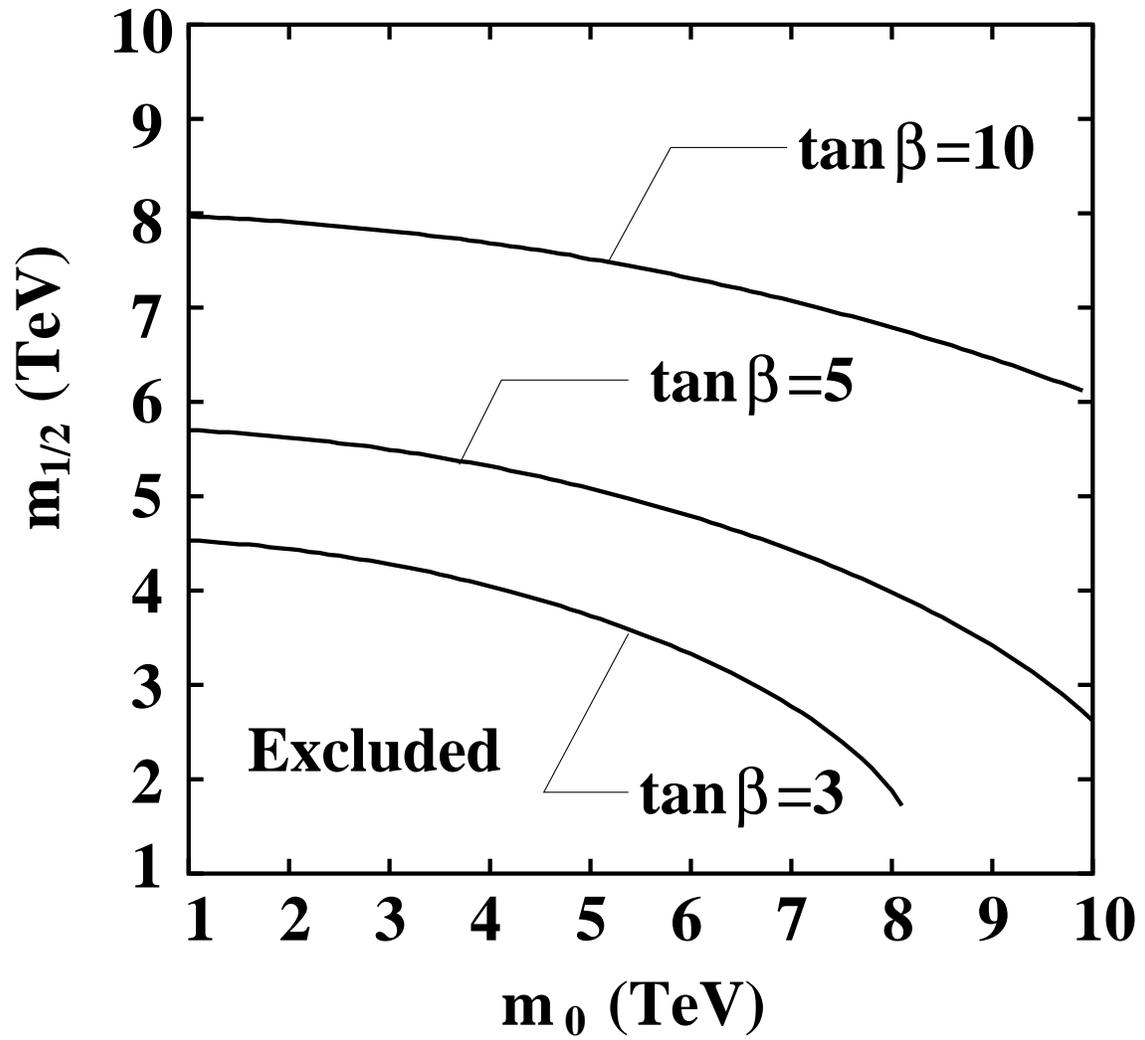}}
% for the preprint server
%\vspace{6cm}
\caption{
Excluded parameter regions restricted by experimental 
neutron EDM's upper limit
}
\end{figure} 

% \begin{table}
% \caption{Please place your table caption here.}
% \begin{tabular}{lrcd} % In second brace, l = left, r = right,
%% c = centered and d = decimal justification.
%% One&Two&Three&Four\\  % Separate items with &. End line with \\
% \tableline % Creates a horizontal line.
% One&Two\tablenote{footnote.}&Three&Four\\ % Place \tablenote{}
%% after item to be footnoted.
% \end{tabular}
% \end{table}

\begin{table}
Table 1. Introduced parameters with mass dimension 
\begin{tabular}{lll}
%\hline\hline
\hline
Symbols  & Energy scales & Descriptions \\
\hline
$M_{P}$ & $10^{19}$(GeV) & 
The Planck mass scale. \\
$M_{SU(5)}$ & $10^{16}$(GeV) & 
The unification scale among the gauge coupling constants. \\
$M_{X}$ & $6.0 \times 10^{10}$(GeV) & 
The spontaneous supersymmetry breaking down scale. \\
$M_{SUSY}$ & 1(TeV) & 
The energy scale emerging the effects of the gauginos. \\
$M_{Z}$ & 91(GeV) & 
The electro-weak scale \\
%\hline\hline
\hline
\end{tabular}
\end{table}


\begin{references}
% Please use the \bibitem command to create references.
\bibitem{He} 
X.-G. He, B. H. J. McKellar, and S. Pakvasa, 
{\em Int. J. Mod. Phys. \/}{\bf A4}(1989), 5011 

\bibitem{Czarnecki} 
A. Czarnecki and B. Krause, 
{\em Phys. Rev. Lett. \/}{\bf 78}(1997), 4339 

\bibitem{Harris} 
P. G. Harris {\em et al.}, 
{\em Phys. Rev. Lett. \/}{\bf 82}(1999), 904 

\bibitem{Romanino} 
A. Romanino and A. Strumia, 
{\em Nucl. Phys. \/}{\bf B490}(1997), 3

\bibitem{KM} 
M. Kobayashi and T. Maskawa, 
{\em Prog. Theor. Phys. \/}{\bf 49}(1973), 652 

\bibitem{Nilles} 
H. P. Nilles and C. G. Switzerland, 
{\em Phys. Rep. \/}{\bf 110}(1984), 1 

\bibitem{Georgi} 
S. Dimopoulos and H. Georgi, 
{\em Nucl. Phys. \/}{\bf B193}(1981), 150 

\bibitem{Guion} 
J. F. Guion and H. E. Haber, 
{\em Nucl. Phys. \/}{\bf B272}(1986), 1

\bibitem{Inui} 
T. Inui, Y. Mimura, N. Sakai and T. Sakai, 
{\em Nucl. Phys. \/}{\bf B449}(1995), 49 

\bibitem{Amaldi} 
U. Amaldi, W. de Boer and H. F\"urstenau, 
{\em Phys. Lett. \/}{\bf B260}(1991), 447 

\bibitem{PDG} 
Particle Data Group, 
http://pdg.lbl.gov/ 

\bibitem{Martin} 
S. P. Martin and P. Ramond, 
{\em Phys. Rev. \/}{\bf D48}(1993), 5365 

\bibitem{Brhlik} 
M. Brhlik, G. J. Good, and G. L. Kane, 
{\em Phys. Rev. \/}{\bf D59}(1999), 115004 

\bibitem{RGE} 
D. J. Casta\~no, E. J. Piard, and P. Ramond, 
{\em Phys. Rev. \/}{\bf D49}(1994), 4882; \\ 
V. Barger, M. S. Berger, and P. Ohmann, 
{\em Phys. Rev. \/}{\bf D49}(1994), 4908; \\ 
L. Alvarez-Gaum\'e, J. Polichinski, and M. B. Wise, 
{\em Nucl. Phys. \/}{\bf B221}(1983), 495; \\ 
L. E. Iba\~nez, 
{\em Nucl. Phys. \/}{\bf B218}(1983), 514. 

\bibitem{SUSY} 
Recent general reviews on the neutron EDM problem: \\
Y. Kizukuri and N. Oshimo, 
{\em Phys. Rev. \/}{\bf D45}(1992), 1806; \\ 
Y. Kizukuri and N. Oshimo,
{\em Phys. Rev. \/}{\bf D46}(1992), 3025; \\ 
R .Garisto and J. D. Wells, 
{\em Phys. Rev. \/}{\bf D55}(1997), 1611; \\ 
T. Ibrahim and P. Nath, 
{\em Phys. Rev. \/}{\bf D57}(1998), 478. 

\bibitem{PD} 
T. Goto and T. Nihei, 
{\em Phys. Rev. \/}{\bf D59}(1999), 115009 
 \end{references}
\end{document}